# OSSE Observations of the Radio Quiet Quasar PG 1416-129

Rüdiger Staubert and Michael Maisack

Institut für Astronomie und Astrophysik der Universität Tübingen, Abteilung Astronomie, 72076 Tübingen, Germany

**Abstract.** We report the detection of the bright radio quiet quasar PG 1416-129 by the OSSE instrument on the Compton Gamma Ray Observatory (CGRO). This object had the hardest spectral index of all known AGN in the energy range 2-20 keV ($\Gamma = 1.1 \pm 0.1$) when observed by Ginga. We find that the source is variable at energies > 50 keV during the OSSE observation, and that the OSSE spectrum is steeper than the Ginga spectrum. The broad-band X-ray spectrum resembles that of Seyfert galaxies, supporting models which interpret radio quiet quasars as the high-luminosity counterparts of Seyferts.

## 1. Introduction

Among radio quiet AGN, Seyfert galaxies and radio quiet QSO are regarded as high- and low-luminosity manifestations of the same phenomenon. This view is based on the multiwavelength spectral energy distributions (e.g. Mas-Hesse et al. 1995) and on relationships between the luminosities of the host galaxy and the nuclear region (e.g. Veron-Cetty and Woltjer 1991). The hard X-ray (>50 keV) spectra of both Seyferts and radio quiet quasars have not been compared so far, since only Seyfert galaxies have been detected at these energies.

We have selected the bright radio quiet quasar PG 1416-129 (z=0.129) as the most promising candidate for a detection above 50 keV by OSSE, due to fact that it had the hardest spectrum of any AGN when observed with Ginga (Williams et al. 1992). The observed photon index in the energy range 2-20 keV was $\Gamma=1.1$. This object is the optically second brightest broad absorption line (BAL) QSO (e.g. Turnshek and Grillmair 1986).

*Send offprint requests to:* R. Staubert; staubert@astro.uni-tuebingen.de

Observations with Ginga and the Compton Gamma Ray Observatory (CGRO) at hard X-rays and $\gamma$-rays have shown that there are mainly two classes of active galactic nuclei (AGN) with characteristic properties at energies above the classical 2-10 keV range: one class comprises the radio quiet Seyfert galaxies, which were found to have X-ray spectra characterised by an underlying power law with a typical photon index of $\Gamma \approx 1.9$ (e.g. Nandra and Pounds 1994), a Compton reflection hump (e.g. White, Lightman and Zdziarski 1988) which peaks at energies 20-50 keV, and a thermal character in the energy range 50 keV to several 100 keV (e.g. Maisack et al. 1993, Johnson et al. 1994). Seyfert galaxies have not been detected at MeV and GeV energies by Comptel or EGRET (Maisack et al. 1995, Lin et al. 1993). Members of the other class, radio loud blazars, were detected in large numbers by EGRET (e.g. Fichtel et al. 1994), some also by Comptel (e.g. Schönfelder 1995) and by OSSE (McNaron-Brown et al. 1995). These objects show power-law spectra at all energy ranges (including X-rays, e.g. Sambruna et al. 1994), indicating a non-thermal origin of the emission. The spectra of blazars are harder than those of Seyferts above 50 keV (McNaron-Brown et al. 1995), with required breaks at several to tens of MeV.

The spectral differences between radio loud and radio quiet AGN are most pronounced either in the radio regime, or at hard X-rays through $\gamma$-rays. This different appearance of these classes is caused by the emission from collimated, relativistic outflows from the core of the AGN, which mainly contribute to the emission in the above energy ranges. Collimated outflows seem to be present only in radio loud AGN. In radio quiet sources, the X-ray emission is dominated by thermal emission from a hot plasma.

## 2. Instrument

OSSE is one of the 4 instruments on CGRO. It is a phoswich-type collimated scintillation counter sensitive between 50 keV and 10 MeV. It consists of four identical detectors which are collimated to a field of view of 3.8 × 11.4 degrees (FWHM). The total effective area at 511 keV is 2000 cm$^2$. Observations are performed by rocking the four detector modules between source and two source-free background fields on either side of the source field along the scan axis in a predefined sequence. A detailed description of the instrument and the data analysis techniques is given in Johnson et al. (1993).

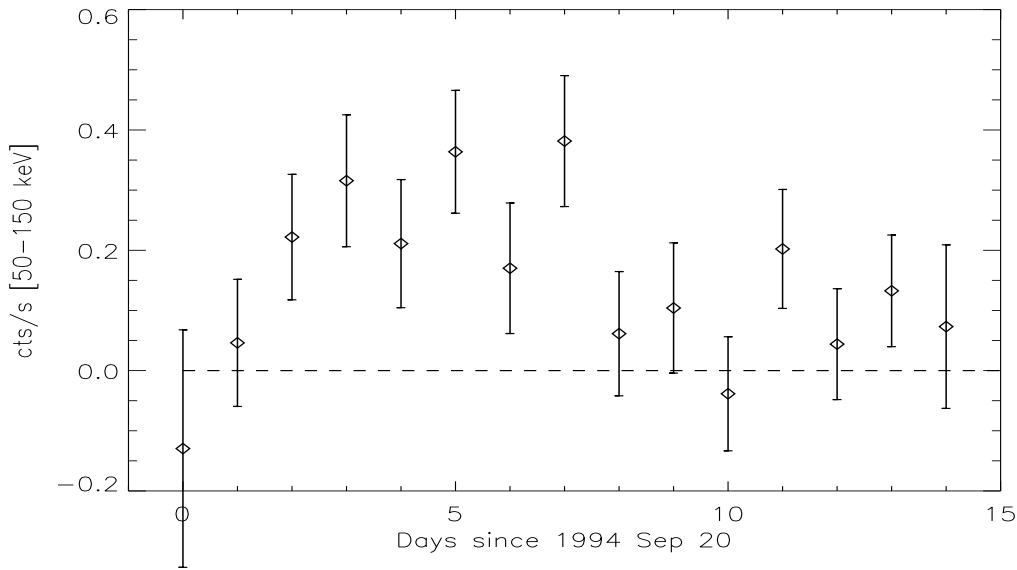

**Fig. 1.** Day to day lightcurve of PG 1416-129 in the energy range 50-150 keV for the sum of 4 detectors

## 3. Observations and Results

PG 1416-129 was observed by OSSE for a duration of two weeks between 1994 Sep 20 and Oct 4. The total on-source observation time with 4 detectors was 361 ksec. The day-to-day lightcurve in the 50-150 keV range for the sum of 4 detectors is shown in Fig 1. It is obvious that the source is variable during the observation. The flux level is higher during parts of the first week of the observation. Adding the data from the period Sep 22-Sep 27, we find a signal with a statistical significance of 4.7$\sigma$. This corresponds to a measured average flux of 6.43$\pm$1.37 $\times$ 10$^{-6}$ photons /(cm$^2$ s keV) in the 50-150 keV range during the time of the observation (144 ksec) (corresponding to 0.27$\pm$0.04 cts/s in Fig.1). Adding all 15 days of observations and all four detectors, no significant signal (at $> 3\sigma$) in the energy range 50-150 keV is detected. The 2$\sigma$ upper limit is 3.94 $\times$ 10$^{-6}$ photons / (cm$^2$ s keV) in the 50-150 keV range (0.17 cts/s). The 2$\sigma$ upper limit on the flux for the later parts of the observation (Sep 28 - Oct 4) is 2.72$\times$10$^{-6}$ photons / (cm$^2$ s keV) in this range (0.11 cts/s). No significant flux is found above 150 keV at any time.

The OSSE data of Sep 22-27 have been fit (see Table 1) by a power law (PL) and by an exponential cut-off which modifies a PL at lower energies (with the fixed Ginga index $\Gamma$= 1.1), (PL$\times$exp.). The PL model gives $\Gamma$= 3.2$\pm$0.5 (the uncertainties throughout the paper correspond to 68% confidence limits for joint variation of parameters). This is considerably steeper than the very hard spectrum found between 2 and 20 keV by Ginga. It is even steeper than the spectra of most Seyferts reported in Johnson et al. (1994) which have an average photon index of $\Gamma$=2.2. In the case of PL$\times$exp. the required normalisation of the PL is considerably higher than the observed Ginga intensities. In Fig. 2 we show the OSSE data together with the published Ginga data (Williams et al. 1992). It can be seen that the 40-100 keV OSSE flux lies above the extrapolation of the (non-simultaneous) Ginga spectrum.

Even though observations by Einstein (Wilkes and Elvis 1987), ROSAT (de Kool and Meurs 1994) and Ginga give no evidence for variability, neither within the individual observations nor between the observations (i.e. less than a factor of 2), the OSSE observations show that the source is clearly variable at higher photon energies. Since we have no way of knowing whether the OSSE and the Ginga observations have seen the source in the same intensity and spectral state or in different ones, we distinguish between those two cases: First, we assume that the (non-simultaneous) data from OSSE and Ginga do in fact belong to the same state, such that combined modelling may be justified. In this case, a spectral bump followed by a cut-off is apparent. Such a spectral shape can be described by different model scenarios. Since the Ginga data are statistically more significant to a degree where a joint fit and the confidence limits are dominated by the Ginga data, we only give representative parameters of trial models. We have attempted a Comptonisation model following Sunyaev and Titarchuk (1980) which can reproduce ($\chi^2_\nu$=1.02 for 25 d.o.f) the combined Ginga and OSSE spectra for a plasma temperature of kT=18 keV and an optical depth for electron scattering of $\tau$=12 (with 68% joint confidence regions of kT=10-30 keV and $\tau$=10-15). Such a spectrum is shown in Fig. 2 as a dashed line. Another approach is to assume that the bump is due to a Compton reflected component, as suggested to be typical for Seyferts by Nandra and Pounds (1994). We have computed Compton reflected spectra using the Green's functions given in White, Lightman and Zdziarski (1988). The solid curve shown in Fig. 2 represents such a spectrum with a reflected component: the photon index of the incident spectrum is $\Gamma$=1.2, the covering factor $f_c$=4[1] ($\chi^2_\nu$=1.28 for 25 d.o.f). The 68% joint confidence regions in this case are $\Gamma$=0.5-1.5 and $f_c$=2-8. The covering factor is quite high compared to values found in many Seyferts (Nandra and Pounds 1994), but is required here to reproduce the sharp cutoff between the Ginga and OSSE spectra. Also, a cutoff in the incident spectrum is required at $\approx$ 150 keV. The existing data cannot discriminate between these two models. Second, we consider the case that the OSSE data may have been associated with a correspondingly higher intensity at 2-20 keV. If the same spectral index as was observed by Ginga ($\Gamma$=1.1) were to apply (to the 2-20 keV range), a power law times exponential cutoff could describe such a spectrum, as given in Table 1. In this case, the broad-band X-ray spectrum of PG 1416-129 is similar to that of Seyfert galaxies, which turn over exponentially between 30-100 keV (Johnson et al. 1994). The best fit e-folding energy for the OSSE data, with $\Gamma$ fixed at 1.1, for this model is 35$\pm$10 keV.

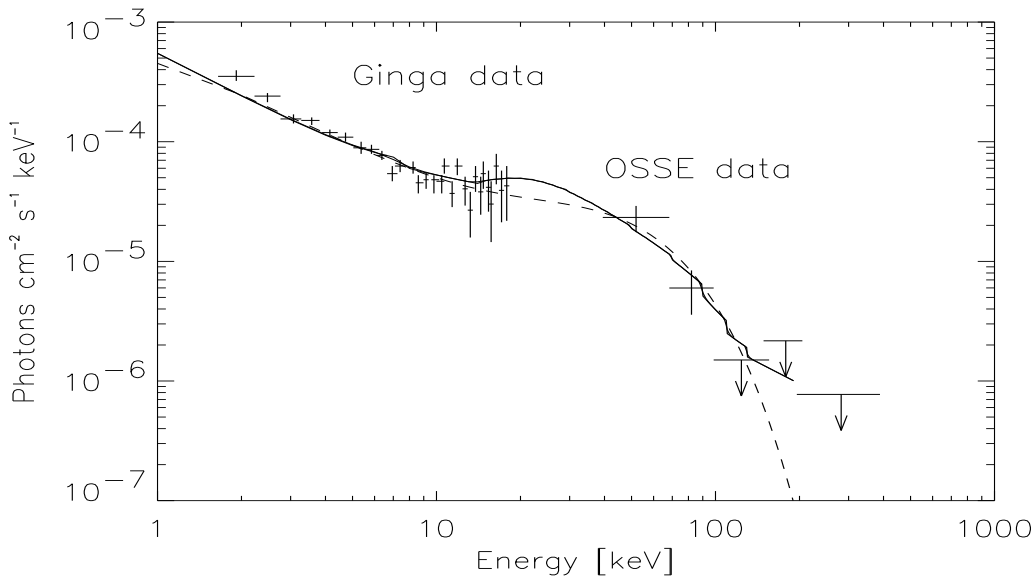

**Fig. 2.** OSSE spectrum of PG 1416-129 for 1994 September 22-27, compared to Ginga data from 1988 (Williams et al. 1992). The dashed curve represents a Comptonised spectrum with kT=18 keV and $\tau$=12. The solid curve represents a spectrum with a Compton reflected component (see text).

**Table 1.** Results of spectral fits to the OSSE data

| Model | $\chi^2_\nu$ | $I^a_{70\ keV}$ | $\Gamma$ | kT [keV] |
|---|---|---|---|---|
| Power Law | 0.81 | 9.21$\pm$1.70 | 3.20$\pm$0.50 | |
| PL $\times$ exp. | 0.86 | 10.70$\pm$1.92 | 1.1[b] | 34.2$\pm$9.4 |

[a] Intensity at 70 keV in $10^{-6}$ photons / (cm$^2$ s keV)
[b] fixed parameter

## 4. Discussion

PG 1416-129 was observed by the ROSAT PSPC in a pointed observation in January 1992. De Kool and Meurs (1994) report a steep spectrum in the 0.1-2.4 keV range, indicative of a soft excess. The spectrum is best described by a power law with photon index 2.2 $\pm$ 0.15, with no intrinsic absorption required. Einstein observations also show a steep spectrum (Wilkes and Elvis 1987) at energies 0.5-3.5 keV, with evidence for a hardening up to 10 keV.

---
[1] the covering is defined as the ratio of reflected and direct component seen by the observer. A factor of $f_c$=1 means that the observer sees both the incident spectrum and the full reflected spectrum as computed in White, Lightman and Zdziarski (1988).

The broad-band spectrum of the radio quiet quasar PG 1416-129, although not observed simultaneously, resembles that of Seyfert galaxies in that it shows evidence for a soft excess, a flat power-law type spectrum between 2 and 20 keV and a steep spectrum above 50 keV. It is significantly different from that of blazars, which show a power law extending from X-rays to MeV energies and a break at several to tens of MeV (McNaron-Brown et al. 1995). The X-ray spectrum of PG 1416-129 has the hardest spectrum observed at 2-20 keV of all AGN and a very steep spectrum above 50 keV, and consequently the most dramatic cutoff between 50 and 100 keV. The Ginga and OSSE data can be modelled by two popular models: Compton reflection or Comptonisation with a low temperature and high optical depth. de Kool and Meurs (1994) have speculated that the hard continuum could be due to a Compton reflected component with a very large covering factor. The steep OSSE spectrum would then represent the high energy flank of the reflection hump. A value of the covering factor $f_c$ significantly higher than unity is required to produce the extreme values of the photon index found by both experiments. Such unusually high values of $f_c$ can be produced assuming a face-on geometry for the accretion disk (Rogers 1991), generally assumed as the reflecting medium. It is not clear under what angle the object is seen. It belongs to the class of Broad Absorption Line (BAL) QSOs, which show signs of outflows at high speed. While outflows from the centre of AGN are generally believed to occur perpendicular to the accretion disk, Begelman et al. (1991) argue in favour of an edge-on geometry for BAL QSOs on the basis of polarisation measurements. Williams et al. (1992) find no evidence for an iron line, a telling characteristic of a Compton reflected spectrum, which would be substantial in this case. The presence of a reflected component thus needs to be addressed by a simultaneous broad-band observation which also provides good energy resolution to determine the intensity and shape of a possible iron line. Compton reflection features are not ubiquitous in Seyferts, however, see for example the *archetypical* Seyfert NGC 4151 (e.g. Maisack and Yaqoob 1991), which is believed to be seen edge-on. We note that NGC 4151 also has a spectrum which turns off more abruptly than that of average Seyfert galaxies (e.g. Zdziarski et al. 1995). PG 1416-129 seems to have an even more dramatic turnover, regardless of the relative normalisation of the Ginga and OSSE spectra. Alternatively to the reflection scenario, the broad-band X-ray spectra of both sources can be described by Comptonisation with low temperatures and high optical depths, contrary to Seyfert galaxies with well-defined substantial reflected components, which have high temperatures and low optical depths.

## 5. Summary

We have detected the first radio quiet quasar, PG 1416-129, at energies >50 keV with OSSE. The source is variable during 14 days of observations. The steep spectrum is similar to that observed from Seyferts in that it has a soft excess, a harder spectrum at 2-20 keV and a steep spectrum above 50 keV. This supports the view that radio quiet quasars are the high luminosity counterparts of Seyfert galaxies. Due to the non-simultaneity of the Ginga and OSSE observations, it is not possible to make a definitive statement about the two possible models for the break between the low- and high-energy spectrum – Compton reflection or thermal Comptonisation – from the exisiting data. While this object needs to be observed simultaneously across the entire X-ray range to constrain better the broad band spectrum, more objects of this class should be observed at these energies (e.g. with OSSE, XTE or SAX) to generalize these conclusions.

*Acknowledgements.* This work was supported by DARA grant 50 OR 92054.